\documentclass[11pt]{article}

\usepackage[
  letterpaper,
  left=1.05in,
  right=1.05in,
  top=0.88in,
  bottom=0.95in,
  headsep=0.24in
]{geometry}
\usepackage[T1]{fontenc}
\usepackage[utf8]{inputenc}
\usepackage{newtxtext,newtxmath}
\usepackage[scaled=0.92]{sourcesanspro}
\usepackage{microtype}

\usepackage{graphicx}
\usepackage{bbm}
\usepackage{amsmath,bm}
\usepackage{array,booktabs}
\usepackage{float}
\usepackage{flafter} 
\usepackage{ragged2e}
\usepackage{enumitem}
\usepackage{caption}
\usepackage{titlesec}
\usepackage{fancyhdr}
\usepackage{xcolor}
\usepackage[normalem]{ulem}
\usepackage[numbers,sort&compress]{natbib}
\usepackage{hyperref}

\definecolor{LinkBlue}{HTML}{264B73}
\definecolor{InkGray}{HTML}{4B5563}
\definecolor{RuleGray}{HTML}{C8CDD3}
\definecolor{qgreen}{rgb}{0,0.45,0}
\definecolor{cutblue}{rgb}{0,0.35,0.8}

\hypersetup{
  colorlinks=true,
  linkcolor=LinkBlue,
  citecolor=LinkBlue,
  urlcolor=LinkBlue,
  pdfauthor={Akiva Goldberg and Nadav M. Shnerb},
  pdftitle={Correlations at criticality in ecological communities}
}
\renewcommand{\thesection}{\Roman{section}}
\renewcommand{\thesubsection}{\Alph{subsection}}

\titleformat{\section}
  {\sffamily\bfseries\large\color{black!88}}
  {\thesection.}{0.65em}{}
\titlespacing*{\section}{0pt}{2.5ex plus .8ex minus .2ex}{1.0ex}

\titleformat{\subsection}
  {\sffamily\bfseries\normalsize\color{black!88}}
  {\thesubsection.}{0.60em}{}
\titlespacing*{\subsection}{0pt}{2.0ex plus .5ex minus .2ex}{0.7ex}

\renewcommand{\headrulewidth}{0.35pt}
\renewcommand{\headrule}{\hbox to\headwidth{\color{RuleGray}\leaders\hrule height \headrulewidth\hfill}}

\newcommand{\rev}[1]{\textcolor{red}{#1}}
\newcommand{\revb}[1]{\textcolor{blue}{#1}}

\newcommand{\onecolumngrid}{}

\newcommand{\preprintincludegraphics}[2][]{%
  \IfFileExists{#2}{\includegraphics[#1]{#2}}{%
    \fbox{\parbox[c][1.65in][c]{0.93\linewidth}{\centering\sffamily\small\color{InkGray}
      Figure file not found:\\[3pt]\texttt{#2}}}%
  }%
}

\setlist{topsep=0.5em,itemsep=0.25em,parsep=0pt}

\begin{document}

\thispagestyle{empty}

\begin{center}
\vspace*{-0.30in}
{\sffamily\bfseries\fontsize{23}{27}\selectfont
Correlations at criticality in ecological communities\par}

\vspace{0.75em}
{\large Akiva Goldberg \quad\textperiodcentered\quad Nadav M. Shnerb\par}
\vspace{0.35em}
{\small\color{InkGray} Department of Physics, Bar-Ilan University, Ramat Gan 52900, Israel\par}
\end{center}

\vspace{0.65em}
\noindent{\color{RuleGray}\rule{\textwidth}{0.6pt}}
\vspace{0.55em}

\begin{center}
\begin{minipage}{0.91\textwidth}
{\sffamily\bfseries\small ABSTRACT}\par\vspace{0.35em}
\small\justifying
Ecological communities are continually reshaped by invasion, exclusion, and diversification, processes that naturally drive them toward the boundary of dynamical stability. Near such a boundary, a soft mode relaxes increasingly slowly and, under stochastic forcing, is expected to dominate the fluctuations, effectively reducing the dynamics to one dimension and generating strong positive and negative abundance correlations. Such correlations have therefore been proposed as signatures of an imminent transition. Here we show that this expectation can fail even arbitrarily close to criticality. The reason is that spectral softness does not guarantee stochastic visibility: the soft mode must receive enough environmental forcing to dominate the fluctuation background generated by the remaining modes. We demonstrate this mechanism in three ecological scenarios: a synthetic feasible community, a local community assembled by immigration from a regional pool, and a community generated by repeated diversification. In all three, communities approach marginal stability without developing the near-perfect pairwise correlations predicted by the single-mode picture. Thus, proximity to ecological criticality need not be visible in equal-time pairwise correlations.
\end{minipage}
\end{center}

\vspace{0.55em}
\noindent{\color{RuleGray}\rule{\textwidth}{0.35pt}}
\vspace{0.55em}

\section{Introduction}

The coexistence of many species and the stability of complex communities are two  of the central puzzles of ecology,
dating back to the classical works of Hutchinson~\cite{hutchinson1961paradox}
and May~\cite{may1972will}. This question is naturally connected to
criticality because ecological communities are not closed systems.
Communities are continually exposed to the addition of new species, either
through immigration from a regional species pool~\cite{kessler2015generalized,Bunin2017,BiroliBuninCammarota2018} or, on longer time scales,
through speciation (or mutations) and the emergence of new  types~\cite{shtilerman2015emergence,araujo2026eco}. Each such
addition tests whether the community can accommodate further diversity
without losing stability or excluding existing species. Sustained assembly
or diversification can therefore drive communities toward the boundaries
that constrain coexistence.

Interest in criticality extends beyond the question of whether
communities can remain stable. Within ecology, proximity to a critical
transition has motivated the search for early-warning signals that could
reveal an impending loss of stability before the transition itself occurs
\cite{Scheffer2009,Chen2019,ChenDNB2012,Dakos2010,Ferreira2025}. More broadly, criticality has attracted considerable
attention in biological physics, where it has been proposed that biological
systems may operate near critical points because such states can confer
functional advantages, including enhanced susceptibility, collective
coordination, and sensitivity to weak perturbations
\cite{MoraBialek2011,Calvo2026}. Criticality is therefore of interest both as a
possible outcome of ecological organization and as a potentially informative
and functionally significant regime in its own right.

Correlation-based statistics provide one experimentally accessible route for
probing proximity to instability. A broad literature has sought signatures
of critical transitions in temporal, spatial, and multivariate correlations
and covariances
\cite{Scheffer2009,Dakos2010,ChenDNB2012,Chen2019,Ferreira2025}.
More recently, Calvo et al.\ focused directly on pairwise covariance in
stochastic many-species communities and showed that their distribution
broadens upon approaching the instability edge, proposing its width as an
estimator of the distance to instability \cite{Calvo2026}.

By and large, the use of such correlations as indicators faces a basic
difficulty: equal-time abundance correlations are not, in general,
intrinsic properties of the ecological dynamics. Recent work has shown that
competition and shared environmental responses contribute to correlations
in opposite directions and may cancel exactly
\cite{camacho2024nonequilibrium,goldberg2026niche,cholsky}.
More generally, the observed correlations are controlled by the mismatch
between the deterministic community dynamics and the covariance structure
of environmental forcing. The same ecological interaction structure can
therefore produce very different correlation patterns under different forms
of environmental variability.

Criticality might nevertheless restore a more universal relation between
dynamics and correlations. Near a stability boundary, the response of the
community becomes strongly amplified along its softest dynamical directions.
Since these directions are selected by the deterministic community dynamics,
one might expect their growing susceptibility to overwhelm the detailed
structure of the stochastic forcing. In this picture, correlations that are
ambiguous far from criticality would become strong and largely determined by
the soft ecological modes close to the transition. Whether this expectation
is actually realized in many-species communities is the question we address
here.

\section{The model: competition and correlated environmental stochasticity} 

To set the scene, let us consider $S$ competing species whose abundances $n_i(t)$ obey
\begin{equation}
\frac{d n_i}{dt}
=
n_i
\left(
r_i-\sum_{j=1}^{S}\alpha_{ij}n_j
\right)
+n_i\eta_i(t),
\end{equation}
where $\alpha=\alpha^{\mathsf T}$ is a symmetric interaction matrix that reflects niche overlap between species~\cite{spaak2020intuitive} such that   $\alpha_{i,i}=1$. Stochasticity is environmental so its amplitude is governed by the abundance~\cite{lande2003stochastic}.

In a realistic system, we expect species' responses to environmental fluctuations to be correlated. We characterize these correlations by the noise correlation matrix,
\begin{equation}
\left\langle
\eta_i(t)\eta_j(t')
\right\rangle
=
\sigma_e^2 C^{(\eta)}_{ij}\delta(t-t'),
\end{equation}
with $
C^{(\eta)}_{ii}=1$. In the absence of more detailed
information about environmental forcing, the minimal biologically
interpretable choice is to separate it into two independent components:
a \emph{matched} component, whose correlations follow the ecological
overlap matrix $\alpha$  and an \emph{idiosyncratic} component acting
independently on each species. We therefore take
\begin{equation}
\left\langle
\boldsymbol{\eta}(t)\boldsymbol{\eta}^{\mathsf T}(t')
\right\rangle
=
\left(
\sigma_{e,m}^2\alpha+\sigma_{e,i}^2 I
\right)\delta(t-t').
\label{eq:mixednoise}
\end{equation}
The first term represents the natural expectation that species with
greater niche overlap also respond more similarly to environmental
variation, either because greater niche overlap reflects reliance on the same resources or, alternatively, because it reflects greater phylogenetic relatedness which, in turn, leads to shared response to environmental variations~\cite{sireci2023environmental}. The second term collects environmental fluctuations specific to
each species. Without additional knowledge of the forcing, there is no natural reason to introduce a more elaborate structure.

\section{Abundance correlations near criticality: negligible or very strong?} 

To analyze the correlations, we assume that the interaction matrix $\alpha$ is positive semidefinite and supports a feasible, stable community. Therefore, all equilibrium abundances are positive and the dynamics are locally stable around the fixed point.

Let $\mathbf n^*$ be a feasible fixed point,
$\alpha\mathbf n^*=\mathbf r$, and define
$N=\mathrm{diag}(n_1^*,\ldots,n_S^*)$.
Linearizing about the fixed point gives the stochastic differential equation,
\begin{equation}
d\delta\mathbf n
=
-N\alpha\,\delta\mathbf n\,dt
+
N\,d\mathbf W ,
\qquad
\left\langle
d\mathbf W\,d\mathbf W^{\mathsf T}
\right\rangle
=
\left(
\sigma_{e,m}^2\alpha+\sigma_{e,i}^2 I
\right)dt .
\label{eq:linearized}
\end{equation}
Introducing $\mathbf y=N^{-1/2}\delta\mathbf n$ and a symmetrized version of the Jacobian, 
\begin{equation}
K=N^{1/2}\alpha N^{1/2},
\end{equation}
this yields, 
\begin{equation} \label{eq6}
d\mathbf y
=
-K\mathbf y\,dt
+
N^{1/2}d\mathbf W,
\end{equation}
with
\begin{equation}
\left\langle
N^{1/2}d\mathbf W\,d\mathbf W^{\mathsf T}N^{1/2}
\right\rangle
=
\left(
\sigma_{e,m}^2 K+\sigma_{e,i}^2 N
\right)dt.
\end{equation}

Note that $K$ is similar to $N\alpha$, so its eigenvalues are the
relaxation rates of the linearized dynamics. $K$ is also a congruence
transformation of the interaction matrix $\alpha$, and Sylvester's law of inertia
guarantees that $K$ is positive definite if and only if $\alpha$ is, so
abundance heterogeneity cannot shift the stability boundary itself. It does,
however, reshape the relaxation spectrum, which can differ substantially
from that of the interaction matrix alone \cite{gibbs2018effect}.

Eq. (\ref{eq6}) facilitates the calculation of abundance correlations. Let
\begin{equation}
\Sigma
=
\left\langle
\delta\mathbf n\,\delta\mathbf n^{\mathsf T}
\right\rangle
\end{equation}
denote the covariance matrix of the abundance fluctuations
$\delta\mathbf n$, and
\begin{equation}
\widetilde{\Sigma}
=
\left\langle
\mathbf y\,\mathbf y^{\mathsf T}
\right\rangle
\end{equation}
the covariance matrix of the rescaled fluctuations
$\mathbf y=N^{-1/2}\delta\mathbf n$. The two are related by
\begin{equation}
\widetilde{\Sigma}
=
N^{-1/2}\Sigma N^{-1/2},
\qquad
\Sigma
=
N^{1/2}\widetilde{\Sigma}N^{1/2}.
\end{equation}
Because this transformation only rescales each species by a positive
constant, the Pearson correlation coefficients are identical in the two
representations:
\begin{equation}
\rho_{ij}
=
\frac{\Sigma_{ij}}{\sqrt{\Sigma_{ii}\Sigma_{jj}}}
=
\frac{\widetilde{\Sigma}_{ij}}
{\sqrt{\widetilde{\Sigma}_{ii}\widetilde{\Sigma}_{jj}}}.
\end{equation}

At this point, two complementary arguments can be made. The first suggests
that abundance correlations should typically remain weak, whereas the second
suggests that this weak-correlation picture may break down as a
critical point is approached.

\begin{figure}[!tbp]
\preprintincludegraphics[width=\linewidth]{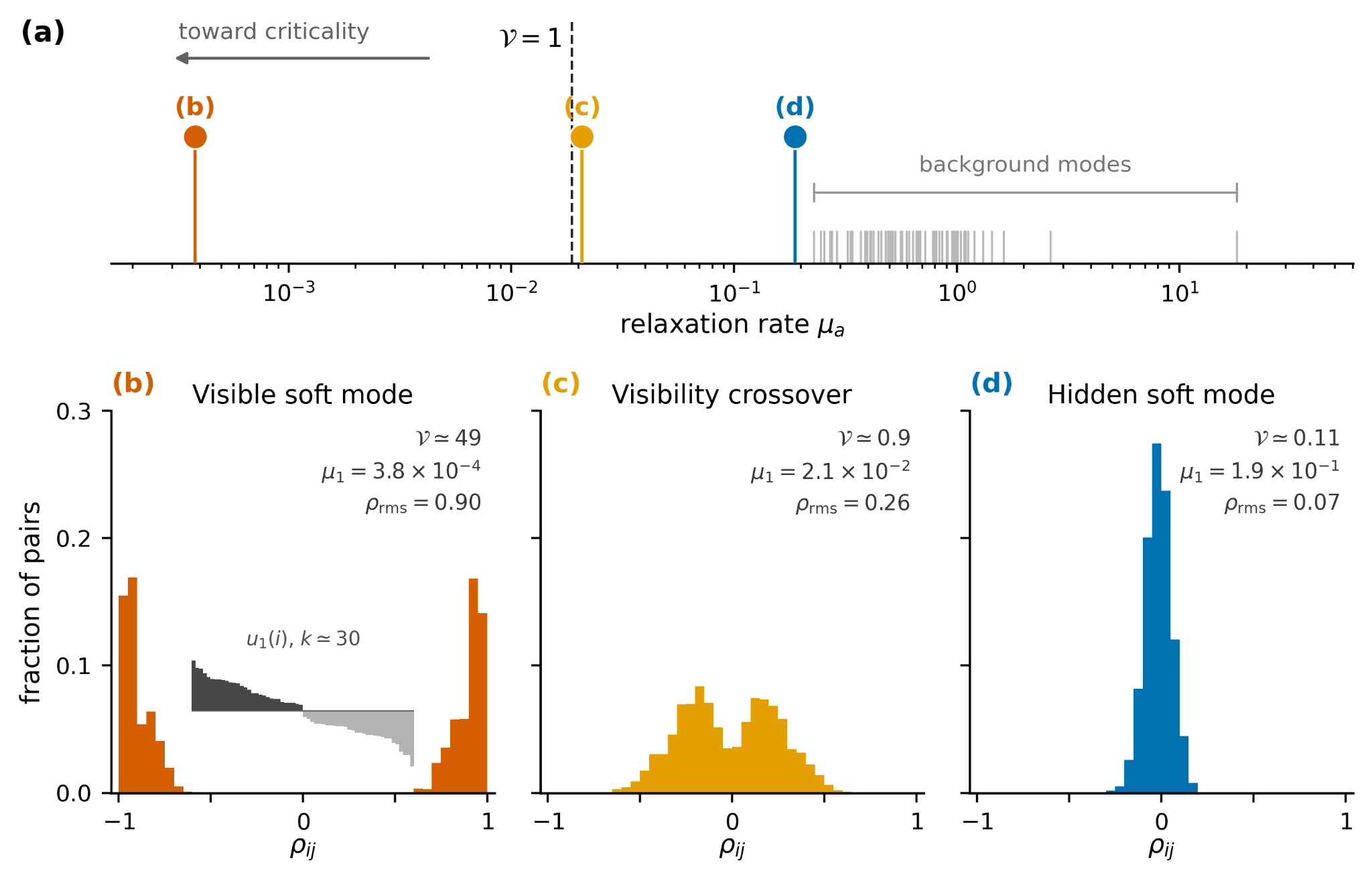}
\caption{\label{fig:naive}
The single-mode expectation in an engineered community.
(a) We construct a feasible competitive community of $S=60$ species with equal
equilibrium abundances and one soft collective mode $u_1$, spread with mixed signs over about half of the community: the inset in panel (b) shows its
weights, and its participation number (the effective
number of species sharing the mode) is $k=30$. All remaining $S-1$ relaxation rates are
held fixed, and only $\mu_1$ is varied. The soft mode becomes visible
in the pairwise statistics once the variance it delivers to the
species carrying it exceeds the variance stored in all other modes;
the ratio of the two is the visibility ${\cal V}$, constructed in
Sec.~\ref{sec:softmodes}, and the dashed line marks the crossover
${\cal V}=1$. 
(b)--(d) Distributions of pairwise Pearson correlations under purely
idiosyncratic forcing for three positions of the soft mode, histograms implement the same bins
and normalization. Below the crossover (d) the soft direction, though
present in the spectrum, is hidden by the fluctuation background and
correlations concentrate near zero. Near the crossover (c) the distribution is broad but shows no accumulation at extreme values. Well beyond the visibility threshold (b) the soft mode dominates every pair it
touches, driving the correlations toward the single-mode limit
$\rho_{ij}=\mathrm{sgn}[u_1(i)u_1(j)]$ and piling them up near
$\rho=\pm1$. This deliberately engineered system provides the reference against which the naturally generated communities are
compared below. Construction details are given in Supplementary
Material~\ref{app:engineered}\revb{, and the numerical values in
Supplementary Material~\ref{C1}}.
}
\end{figure}

\begin{enumerate}

\item In the rescaled variables the stationary covariance
$\widetilde\Sigma$ satisfies
\begin{equation}
K\widetilde\Sigma+\widetilde\Sigma K
=
\sigma_{e,m}^2 K+\sigma_{e,i}^2 N.
\label{eq:LyapMixed}
\end{equation}
Let $U=(\mathbf u_1,\ldots,\mathbf u_S)$ diagonalize $K$,
\[
U^{\mathsf T}KU=\mathrm{diag}(\mu_1,\ldots,\mu_S).
\]
We denote quantities expressed in this eigenmode basis by a prime, defining
\[
\widetilde\Sigma'=U^{\mathsf T}\widetilde\Sigma U,
\qquad
N'=U^{\mathsf T}NU.
\]
Thus, $a$ and $b$ label the eigenmodes of $K$. In this basis,
Eq.~(\ref{eq:LyapMixed}) can be solved element by element, yielding
\begin{equation}
\widetilde\Sigma'_{ab}
=
\frac{\sigma_{e,m}^2}{2}\delta_{ab}
+
\sigma_{e,i}^2
\frac{N'_{ab}}{\mu_a+\mu_b}.
\label{eq:SigmaModes}
\end{equation}
Transforming back to species space gives
\begin{equation}
\widetilde\Sigma_{ij}
=
\frac{\sigma_{e,m}^2}{2}\delta_{ij}
+
\sigma_{e,i}^2
\sum_{a,b}
u_a(i)u_b(j)
\frac{N'_{ab}}{\mu_a+\mu_b}.
\label{eq:SigmaSpecies}
\end{equation}

Importantly, the matched forcing produces an isotropic covariance and
therefore no interspecific correlations. 

We emphasize that the vanishing of correlations at the matched point is not
an artifact of the linearized analysis. It can be shown exactly, for arbitrary
noise strength, that matched environmental forcing produces no equal-time
abundance covariance between distinct species. Thus, even arbitrarily close
to criticality, perfectly matched forcing by itself generates no interspecific
correlations.

Any off-diagonal covariance must therefore originate from the mismatch between
the stochastic forcing and the ecological interactions. In the present
decomposition, this contribution is generated entirely by the idiosyncratic
component and is linear in $\sigma_{e,i}^2$. The diagonal variances, by
contrast, also receive the matched contribution, which grows with
$\sigma_{e,m}^2$. Since correlations normalize the off-diagonal
covariance by the variances of the two species, this observation suggests a
first, simple expectation: in the presence of a substantial matched component
of environmental variability, interspecific abundance correlations should
typically remain weak.

\item This weak-correlation picture need not persist close to
the stability boundary. Let
\begin{equation}
K\mathbf u_a=\mu_a\mathbf u_a,
\end{equation}
with loss of stability occurring as the smallest relaxation rate
$\mu_1$ approaches zero. From Eq.~(\ref{eq:SigmaModes}), the variance
carried by the softest mode is
\begin{equation}
\widetilde\Sigma'_{11}
=
\frac{\sigma_{e,m}^2}{2}
+
\frac{\sigma_{e,i}^2}{2}
\frac{\nu_1}{\mu_1},
\qquad
\nu_1\equiv \mathbf u_1^{\mathsf T}N\mathbf u_1 .
\label{eq:softvariance}
\end{equation}
Thus, for any nonzero idiosyncratic component, one might expect the
soft-mode contribution to grow without bound as $\mu_1\to0$, eventually
overwhelming both the matched contribution and the fluctuations carried
by the remaining modes.

If a single soft mode indeed dominates the fluctuations, then $\mathbf y(t)$ is proportional to $u_1$  and the motion of the community becomes effectively one-dimensional.
Its projection onto any pair of species with nonzero weight on the mode
is therefore a straight line,
\begin{equation}
y_j(t)\simeq
\frac{u_1(j)}{u_1(i)}\,y_i(t),
\end{equation}
so that
\begin{equation} \label{pm1}
\rho_{ij}\rightarrow
\operatorname{sgn}\!\left[u_1(i)u_1(j)\right].
\end{equation}
This argument therefore predicts a qualitative change near criticality.
The weak, mismatch-dependent correlations expected away from
the stability boundary should give way to correlations concentrated near
$\pm1$, producing the U-shaped distribution shown in
Fig.~\ref{fig:naive}. In this picture, any departure from perfectly
matched forcing is eventually amplified by the diverging susceptibility,
the detailed structure of the mismatch becomes secondary, and the
geometry of the critical mode determines the observed correlations.
\end{enumerate}

Both arguments are correct, but they apply to different aspects of the
same limit. Close to the stability boundary
the susceptibility of a soft mode can strongly amplify even a small
mismatch. However, whether this amplification actually drives pairwise
correlations toward $\pm1$ depends on how the soft mode itself behaves as
$\mu_1\to0$.  We examine this singular limit more
carefully in the next section.

\section{The effect of soft modes on correlations}  \label{sec:softmodes}

Let us now examine more carefully the limit $\mu_1\to0$. We first consider
the simplest case in which a single, well-separated mode becomes soft while
the remaining relaxation rates stay finite. For this mode, define its
exposure to idiosyncratic forcing,
\begin{equation}
\nu_1
\equiv
N'_{11}
=
\mathbf u_1^{\mathsf T}N\mathbf u_1
=
\sum_i n_i^*u_1(i)^2 .
\label{eq:nu}
\end{equation}
The $(1,1)$ component of Eq.~(\ref{eq:SigmaModes}) then contributes the
rank-one term
\begin{equation}
A_1\,\mathbf u_1\mathbf u_1^{\mathsf T},
\qquad
A_1\equiv
\frac{\sigma_{e,i}^2\nu_1}{2\mu_1},
\label{eq:A1}
\end{equation}
to the covariance in species space.

We collect all remaining contributions into a background matrix $B$,
writing
\begin{equation}
\widetilde\Sigma
=
A_1\,\mathbf u_1\mathbf u_1^{\mathsf T}
+
B.
\label{eq:signalbackground}
\end{equation}
Here $B$ contains both the isotropic covariance generated by the matched
forcing and all contributions other than the $(1,1)$ soft-mode term from
the idiosyncratically driven covariance. The correlation between species
$i$ and $j$ can therefore be written as
\begin{equation}
\rho_{ij}
=
\frac{
A_1u_1(i)u_1(j)+B_{ij}
}{
\sqrt{
\left[A_1u_1(i)^2+B_{ii}\right]
\left[A_1u_1(j)^2+B_{jj}\right]
}
}.
\label{eq:rhoSignalBackground}
\end{equation}

Equation~(\ref{eq:rhoSignalBackground}) makes transparent the practical requirement
for the appearance of strong, $\pm1$ correlations. For a particular pair
of species to approach perfect correlation, the soft-mode contribution
must dominate the background for both species,
\begin{equation}
A_1u_1(i)^2 \gg |B_{ii}|,
\qquad
A_1u_1(j)^2 \gg |B_{jj}|,
\label{eq:pairvisibility}
\end{equation}
with the background covariance $B_{ij}$ likewise negligible compared with
$A_1u_1(i)u_1(j)$. In that limit Eq.~(\ref{eq:rhoSignalBackground})
reduces to the perfect-correlation result derived above, Eq. (\ref{pm1}).

Thus, the divergence of $1/\mu_1$ is not by itself enough to guarantee
visible critical correlations. Whether the soft mode dominates a given
pair depends on three quantities: how strongly the mode is excited through
$A_1$, how much weight it places on each of the two species, and how large
the remaining fluctuation background is. We now examine these requirements
in turn.

\subsection{Localization of the soft mode}

The first requirement for a visible critical signal concerns how the soft
mode is distributed across species. If the mode is effectively supported on
$k$ species, its typical weight on a participating species is of order $u_1(i)^2\sim \frac{1}{k}.$

It is therefore useful, as suggested by Suweis et
al.~\cite{suweis2015effect},  to characterize its localization by the participation
number
\begin{equation}
k\equiv
\frac{1}{\sum_i u_1(i)^4}.
\label{eq:participation}
\end{equation}
For species within the support of the mode, the soft
contribution to the variance is then typically of order $A_1/k$.
Suweis et
al.~\cite{suweis2015effect} found localization associated with degree
heterogeneity and concentrated on well-connected species. Here, as we show
below, the mechanism behind localization is abundance heterogeneity itself, which causes the soft modes to be localized on rare species.

Localization  has two
opposite consequences. A more localized mode places more weight on each
species that carries it and can therefore more easily dominate the local
background. At the same time, it affects only a limited fraction of the
community: a mode supported on $k$ species can generate strong correlations
for only $O(k^2)$ pairs out of the $O(S^2)$ possible pairs. Hence a localized
critical mode may produce a few strongly correlated pairs without generating
a community-wide accumulation of correlations near $\pm1$.

In the extreme case $k\to1$ the mode is confined to a single species
and correlates no pair at all. The quantity relevant to pair statistics is
therefore not necessarily $\mu_1$, but the relaxation rate $\mu_s$ of the
softest mode that is not so confined (Supplementary
Material~\ref{app:models}). The two differ in about half of the communities
studied below, and $\mu_s$ is used throughout.

In ecological communities, localization of soft modes can arise naturally
from abundance heterogeneity. Recall that
\begin{equation}
K_{ii}=n_i^*\alpha_{ii},
\qquad
K_{ij}=\sqrt{n_i^*n_j^*}\,\alpha_{ij}.
\end{equation}
With the normalization $\alpha_{ii}=1$, the diagonal element associated with
a rare species is simply $K_{ii}=n_i^*$, while all couplings connecting that
species to the rest of the community are suppressed as $\sqrt{n_i^*}$.

This suggests a useful analogy with a Schr\"odinger operator on the species
network: a rare species has both a low on-site energy and weak hopping to its neighbors. In the weak-coupling limit, the eigenmodes approach individual-species directions and the relaxation rates reduce to the known abundance-tracking result,
$\lambda_i \simeq -n_i^*$ \cite{stone2018feasibility}. Our result extends this picture beyond the
weak-coupling limit: rarity-induced localization persists even when
interactions substantially mix the species, producing soft modes analogous
to localized low-lying band-edge states in disordered systems. Abundance
heterogeneity therefore provides a natural route to soft modes localized on
rare species.

Rarity, however, also affects how strongly such a mode is driven. Its exposure
to idiosyncratic forcing is
\begin{equation}
\nu_1
=
\sum_i n_i^*u_1(i)^2 .
\end{equation}
If the soft mode is localized on a rare species of abundance
$n_i^*=\epsilon\ll1$, its exposure to the idiosyncratic forcing scales with
that abundance, $\nu_1\simeq\epsilon$. Thus the same abundance heterogeneity
that makes the mode soft also suppresses the forcing projected onto it. If
the softness itself is generated by rarity, so that $\mu_1\propto\epsilon$,
then 
\begin{equation}
\frac{\nu_1}{\mu_1}={\cal O}(1),
\end{equation}
and the soft-mode amplitude
\begin{equation}
A_1=
\frac{\sigma_{e,i}^2}{2}\frac{\nu_1}{\mu_1}
\end{equation}
need not diverge as $\mu_1\to0$. A community can therefore become spectrally
critical through a rare, localized species while the corresponding stochastic
mode remains only weakly amplified.

There are two further consequences of rarity-induced localization.
First, in the extreme case $k \approx 1$, the soft mode is confined to a single
species and cannot by itself generate any interspecific correlation;
correlations with other species can arise only through the small tails of
the eigenvector. Second, because the mode is concentrated on rare species,
its contribution to absolute community-level fluctuations is itself small.

Localization has one further consequence for pairwise covariance. Suppose the soft mode is localized on two species, $i$ and $j$, and the components $u_1(i)$ and $u_1(j)$ are both positive, so the  soft mode contributes to positive covariance between these species. However, in all other background modes $u_1(i)$ and $u_1(j)$ must have opposite sign due to orthogonality, and therefore the background contribution to the covariance and to the correlations reduces the contribution of the soft mode.

\subsection{Conditions for strong correlations}

We can now summarize the conditions under which a soft mode produces strong
pairwise correlations. Consider first purely idiosyncratic forcing,
$\sigma_{e,m}=0$. If the soft mode is effectively distributed over $k$
species, its contribution to the variance of a species in its support is
typically
\begin{equation}
\frac{\sigma_{e,i}^2}{2}
\frac{\nu_1}{\mu_1 k}.
\label{eq:softperspecies}
\end{equation}
The remaining modes provide a fluctuation background. For a scalar estimate of the background, we average its diagonal
contribution over species. The mixed modal terms then vanish by
orthogonality, giving
\begin{equation}
\frac{1}{S}\sum_i B_{ii}
=
\frac{\sigma_{e,i}^2}{2S}
\sum_{a\neq1}\frac{\nu_a}{\mu_a}.
\label{eq:meanbackground}
\end{equation}
Thus, strong correlations between two species carrying typical weight on
the soft mode require
\begin{equation}
 \frac{\nu_1}{\mu_1 k}
\gg {\cal B} \equiv 
\frac{1}{S}
\sum_{a\neq1}\frac{\nu_a}{\mu_a}.
\label{eq:visibilitycondition}
\end{equation}

This condition makes the three requirements transparent. The decrease of
$\mu_1$ must generate a genuine amplification of $\nu_1/\mu_1$, rather than
being compensated by a simultaneous decrease in the exposure $\nu_1$.
The mode must also place sufficient weight on the species being observed,
so that its signal is not diluted over too many species. Finally, this signal
must exceed the fluctuation background generated by all remaining modes.

For two species with comparable soft-mode weights, we return to
Eq.~(\ref{eq:rhoSignalBackground}). If the remaining modes are sufficiently
delocalized, their signed contributions to the off-diagonal background
$B_{ij}$ largely cancel, while their contributions to the diagonal variances
add. If, as a first estimate, we neglect this compensating off-diagonal
background and replace the diagonal background by its species average,
Eq.~(\ref{eq:rhoSignalBackground}) gives
\begin{equation}
|\rho_{ij}|
\sim
\frac{
\nu_1/(\mu_1 k)
}{
\nu_1/(\mu_1 k)
+
{\cal B}
}.
\label{eq:rhoEstimate}
\end{equation}
This estimate is systematically optimistic because the orthogonality
constraint discussed above also reduces the covariance in the numerator.
In our numerical experiments (presented below)  we find that the remaining
background contributions cancel approximately one half of the soft-mode
covariance. Accordingly, the measured
correlations are approximately a factor of two smaller than the estimate in
Eq.~(\ref{eq:rhoEstimate}).

Eq.~(\ref{eq:rhoEstimate}) implies that correlations approach the $\pm1$ limit derived above only when
the soft-mode contribution per participating species becomes much larger
than the background ${\cal B}$. If the remaining modes have typical relaxation rate
$\mu_{\rm typ}$ and comparable $O(1)$ exposures, ${\cal B}$ is of order
$1/\mu_{\rm typ}$, rather than $S/\mu_{\rm typ}$, since the contributions of
the $S$ modes are each spread over $S$ species.

A matched component of the forcing only strengthens this requirement: it
adds $\sigma_{e,m}^2/2$ to the variance of every species without generating
off-diagonal covariance. Strong critical correlations therefore require not
merely a small $\mu_1$, but a soft mode whose amplification survives the
loss of noise exposure and is large enough, on the species that carry it,
to dominate all other sources of variance.

The comparison above defines a central quantity of this paper. We
call
\begin{equation}
{\cal V}
\;\equiv\;
\frac{\nu_1}{\mu_1\,k\,{\cal B}}
\label{eq:visibility}
\end{equation}
the \emph{visibility} of the soft mode: the ratio between the
variance the mode delivers to a typical species carrying it,
$\nu_1/(\mu_1 k)$, and the variance ${\cal B}$ stored per species in
all remaining modes. The naive expectation of correlations near $\pm1$
corresponds to ${\cal V}\to\infty$, and the crossover demonstrated in  
 Fig.~\ref{fig:naive} sits at
${\cal V}=1$. This criterion is deliberately permissive: we ask whether criticality becomes visible even in the pairs most favorably positioned to reveal the soft mode.

The analysis above establishes an asymptotic point: spectral proximity to instability alone does not guarantee visible pairwise correlations, because softening may be offset by loss of noise exposure or rendered ineffective by localization. We now ask whether these obstructions arise in practice along plausible ecological routes that generate near-marginal communities.

\section{Natural routes to criticality } \label{sec:twoarguments}

We now ask how this visibility budget is realized in communities generated
by increasingly endogenous forms of organization. We compare feasible
synthetic communities, local communities assembled by immigration, and
communities generated by repeated diversification. For the synthetic community we have chosen the parameters (richness, mean niche overlap, heterogeneity) such that the resulting community tends to be either unfeasible or almost critical. In the case of the local community, the parameters were chosen for the
regional pool, in the multiple-equilibria regime~\cite{Bunin2017}, and the structure of the local community emerges
from the assembly process. The evolutionary model allows the community to diversify until it hits the critical point. Full protocols and parameters for the three models are given in Supplementary Material~\ref{app:models}.

\begin{figure}[!tbp]
\preprintincludegraphics[width=\textwidth]{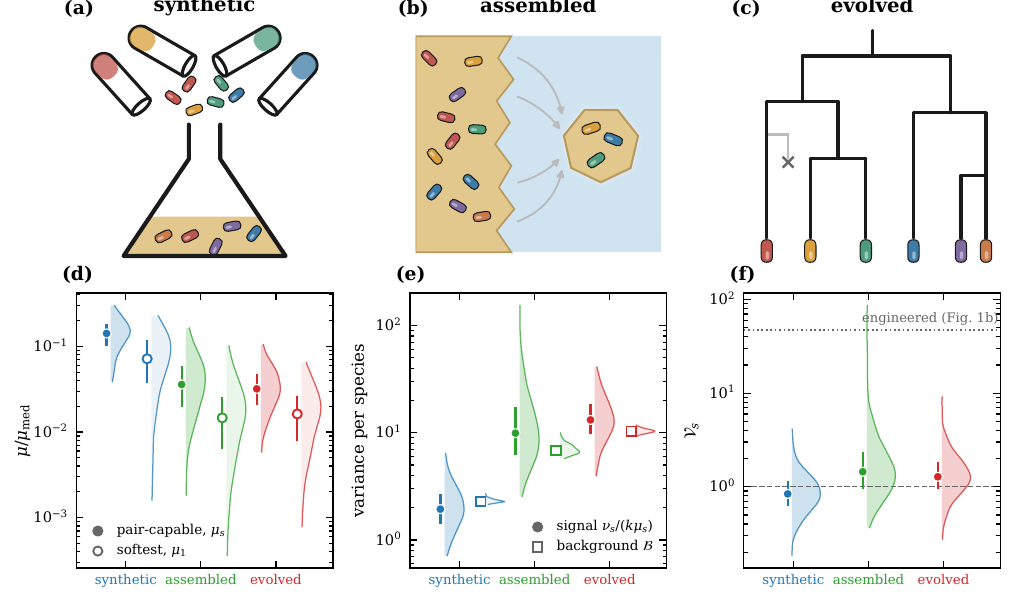}
\caption{\label{fig:scenarios}
Three natural routes to criticality and their visibility budgets.
(a)--(c) Schematic representation of the three generative mechanisms:
a synthetic feasible community drawn from unrelated species, a local
community assembled by immigration from a regional pool, and an evolved
community generated by repeated in situ diversification and ecological
sorting. 
(d) Distributions of normalized relaxation rates $\mu_s/\mu_{\rm med}$ (filled) and
$\mu_1/\mu_{\rm med}$ (open). Both organized ensembles lie  below the synthetic one. (e) The variance delivered per participating species by the soft mode,
$\nu_s/(k\mu_s)$ (filled circles), and the background variance per species,
${\cal B}=S^{-1}\sum_{a\neq s}\nu_a/\mu_a$ (open squares), for purely
idiosyncratic forcing. Signal and background grow together. (f) The visibility ratio ${\cal V}_s=\nu_s/(k\mu_s{\cal B})$, with the crossover
${\cal V}=1$ (dashed). The corresponding value for the  engineered reference of
Fig.~\ref{fig:naive}(b), is dotted. For all three ensembles the median stays of
order unity: the mechanisms that bring the community
closer to criticality simultaneously soften the fluctuation background
against which the critical mode must be observed. Numerical values are
collected in Supplementary Material~\ref{C2}.}
\end{figure}

The three mechanisms do not bring the community equally close to
instability. The normalized relaxation rate of the pair-capable mode
drops  from the synthetic ensemble to either
organized one [Fig.~\ref{fig:scenarios}(d)]. Thus,
greater endogenous organization is accompanied by a deeper
approach to marginality: assembly can select a compatible subset of the
regional pool, while diversification can also build structure into the
interaction matrix itself.

The increased spectral depth is indeed transmitted to the soft mode: the
decrease of \rev{$\mu_s$} is not offset by a parallel loss of exposure, so the
excitation \rev{$A_s=\nu_s/\mu_s$} grows as \rev{$\mu_s$} decreases. It follows the same
ordering, largest in the evolved communities and smallest in the synthetic
ones. Crucially, however, the fluctuation background 
does the same. The quantity ${\cal B}$ increases together with $A$
[Fig.~\ref{fig:scenarios}(e)]. Thus the mechanisms that allow a community to approach
marginality do not produce an isolated soft direction. They soften a broader
part of the fluctuation spectrum, increasing the variance against which the
softest mode must be observed. In the synthetic ensemble both quantities remain small, while assembly
and diversification enhance both by comparable factors, the evolved
communities carrying the largest amplification together with the largest
background, reflecting their dense band of near-soft modes.

These two trends nearly compensate. Despite the large hierarchy in
$\mu_s/\mu_{\rm med}$ and in the absolute soft-mode excitation, the
median visibility ratio remains of order unity and varies little among the three
ensembles [Fig.~\ref{fig:scenarios}(f)]. Increasing self-organization therefore makes the
community more susceptible in absolute terms, but does not make its softest
mode more dominant relative to the fluctuations carried by the rest of the
community. Self-organization buys spectral depth at the price of spectral background.

Equation~(\ref{eq:rhoEstimate}) converts this variance budget  into a naive, variance-only estimate of the pairwise correlation by neglecting the off-diagonal
background $B_{ij}$. It correctly predicts that correlations should
remain far from unity but, as we show below [Fig.~3(b)], systematically
overestimates their magnitude, for a reason anticipated by the
orthogonality constraint discussed above.

\begin{figure}[!tbp]
\preprintincludegraphics[width=\linewidth]{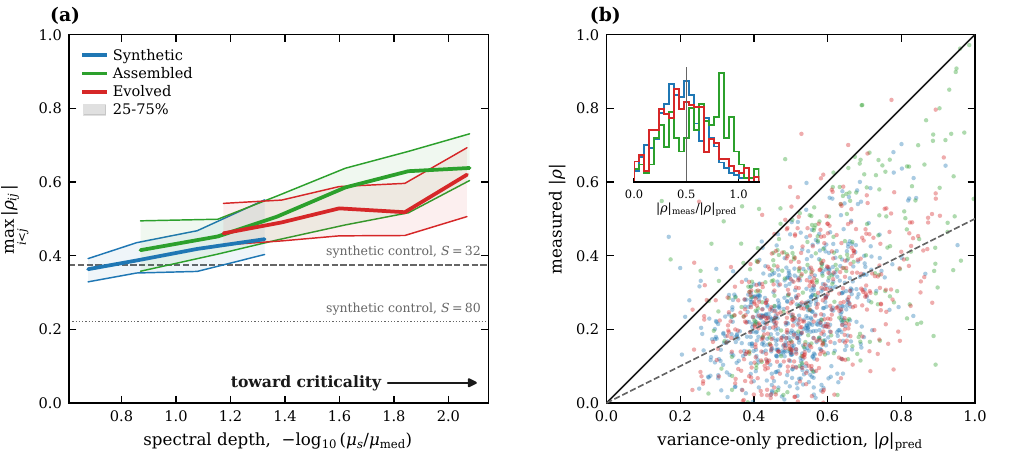}
\caption{\label{fig:corrs}
Pairwise correlations in the three natural ensembles. (a) The strongest pairwise correlation in each community,
$\max_{i<j}|\rho_{ij}|$, against the spectral depth of the pair-capable
mode, $-\log_{10}(\mu_s/\mu_{\rm med})$. Heavy lines are binned medians, thin lines and the faint
wash the interquartile range. Synthetic, assembled and evolved communities
are blue, green and red, and horizontal dashed lines are size-matched synthetic
controls. The strongest correlation grows toward the edge in every
ensemble, but levels off near $0.6$, far below the single-mode limit
$|\rho|=1$.
(b) The orthogonality tax. Measured correlations of the best pair against
the variance-only estimate of Eq.~(\ref{eq:rhoEstimate}), which neglects
the off-diagonal background $B_{ij}$. Solid line, equality; dashed line, a
reduction by a factor of two; inset, the distribution of the ratio.
Numerical values and the matched-depth comparison with the controls are
given in Supplementary Material~\ref{C3}, and the mode-by-mode origin of
the factor of two in Supplementary Material~\ref{C4}.}
\end{figure}

Figure~\ref{fig:corrs} shows the correlations themselves across the
three natural ensembles. Organization does enhance the strongest
pairwise correlations: at matched spectral depth, assembled and evolved
communities lie systematically above synthetic controls of the same size
(Fig.~\ref{fig:corrs}(a)). Within each ensemble the strongest correlation
also grows with spectral depth. The growth is nevertheless
bounded: the median levels off near $0.6$ at the deepest bins.

It should be stressed that the quantity plotted is the single strongest
pair in a community, an extreme-value statistic, and not the typical
correlation. The distinction matters, because the signature of the
single-mode limit is not one strong pair but a collective one (like the U-shaped curve in  Fig.~\ref{fig:naive}). In the strongest outlier community only  $0.08$ of the correlations were above  $0.9$.  Proximity to criticality is therefore necessary but
not sufficient for strong pairwise correlations; the missing ingredient is
the visibility of the soft mode.

The correlations are in fact weaker than the variance-only estimate of
Eq.~(\ref{eq:rhoEstimate}). As shown in Fig.~\ref{fig:corrs}(b), the
measured values are systematically smaller by approximately a factor of
two. This reduction is the orthogonality effect discussed above: for
pairs carrying large weight on the localized soft mode, the
mode-diagonal contribution of the remaining spectrum has predominantly
the opposite sign and cancels roughly half of the soft-mode covariance,
while cross-mode contributions are small.

\section{Discussion}

Our results separate proximity to instability from its visibility in
pairwise abundance correlations. A small relaxation rate $\mu_1$ provides
a large dynamical susceptibility, but this gain alone does not determine
the observed correlations. The soft direction must also remain exposed to
environmental forcing, carry appreciable weight on the species being
observed, and dominate the fluctuations stored in the rest of the spectrum.
These conditions are simultaneously realized in the deliberately engineered
single-mode community of Fig.~\ref{fig:naive}, but not in the three natural
routes to marginality studied here.

The microscopic obstruction is different in each route, although the
outcome is the same. In synthetic communities, softness is typically tied
to rarity, which simultaneously reduces the forcing projected onto the
soft mode. Assembly in local communities can coordinate abundance and interaction structure and
thereby strengthen the soft direction~\cite{kessler2025interaction}, while diversification can generate
still deeper collective modes; however, the latter arise together with a
band of near-soft directions that raises the fluctuation background.
Localization adds a further, more general constraint: when a soft mode
places large weight on a small set of species, orthogonality forces the
remaining spectrum to carry compensating pair weight on those same species.
The background can therefore oppose the soft-mode covariance as well as
inflate the variances against which it is normalized. Thus the failure of
the single-mode picture does not originate from a single mechanism, but
from different structural constraints associated with the ways ecological
communities approach marginality.

This does not mean, however, that ecological organization leaves no
signature in pairwise correlations. As shown in Fig.~\ref{fig:corrs}(a),
assembled and evolved communities develop stronger pair correlations than synthetic
controls of comparable size. What fails is
the use of these correlations as a direct measure of spectral distance:
although the strongest correlation does increase with
spectral depth, it saturates and the
near $\pm1$ signature of the isolated soft-mode limit does not emerge.
A measured value of $|\rho|$ therefore cannot be inverted into a distance
to the edge without knowing the architecture and the size of the
community.

The relevant distinction is therefore between correlations as signatures
of \emph{endogenous organization} and correlations as indicators of \emph{distance to
criticality}. The former can remain informative even when the latter
relationship is weak. In particular, assembly biases the strongest
correlations toward rare species, whereas diversification leaves a
different fingerprint, an anomalously large fluctuation amplitude relative
to the spectral depth associated with its inherited band of near-soft
modes. That assembled interaction matrices carry structure beyond their entry
statistics is already established, see~\cite{kessler2025interaction, shtilerman2015emergence}.

This raises the question of how our results relate to existing multivariate signatures of critical transitions. In the classical codimension-one setting, a single critical direction separates from the rest of the spectrum and comes to dominate the observed fluctuations. The engineered
community of Fig.~\ref{fig:naive} realizes precisely this limit: the
isolated soft mode becomes strongly excited, the leading covariance
eigenvalue becomes dominant, and strong pairwise correlations
emerge, as anticipated by covariance-based early-warning approaches
\cite{Chen2019} and by the
dynamical-network-biomarker methods \cite{ChenDNB2012}. In these two works, however, the stochastic forcing is introduced in a form
that does not depend on population size (injected directly into the
mode with fixed amplitude \cite{ChenDNB2012}, or applied
additively in \cite{Chen2019}) so the
exposure of the soft mode is fixed by assumption. Under such forcing, localization does not reduce visibility. Ecological communities violate
this premise, because environmental forcing scales with abundance
and the softness itself originates in abundance heterogeneity.
Rarity can then soften a mode while simultaneously suppressing the
forcing projected onto it, whereas diversification generates a band
of comparably slow and excited directions rather than a single
dominant one. Thus the indicators themselves behave as expected in
the single-mode limit; what fails in the natural communities is the
emergence of that limit as a generic consequence of approaching
marginality.

The failure of the single-mode limit does not, however, erase the
signature of marginality from the covariance altogether.
Random-matrix calculations of Ornstein--Uhlenbeck ensembles find a
power-law tail in the covariance spectral density at marginal
stability \cite{Ferreira2025}. Like the early-warning constructions
above, this rests on protected exposure, here enforced by Onsager
reversibility, under which detailed balance ties each mode's
forcing to its relaxation rate. Our communities obey no such
constraint, yet along the evolutionary route the soft-band
covariance grows with no loss of exposure
(Fig.~\ref{fig:scenarios}e). An enhanced spectral tail thus remains a
viable signature of marginality, at least along this route, even
though the background keeps it out of the pairwise correlations.

A complementary covariance-based approach was proposed in Ref.~\cite{Calvo2026}, where the width of pairwise associations increases as instability is approached. We find a closely related broadening of the Pearson-correlation distribution. However, our results (Supplementary Material~\ref{app:numbers},
Fig.~\ref{fig:rms}) show that such a width can serve as a proxy for distance to instability only within comparable community architectures: at matched size and spectral depth, different architectures can produce systematically different correlation widths.

There is also an important mechanistic distinction. The construction of Ref.~\cite{Calvo2026} follows an engineered route to instability, analogous to the engineered community in Fig.~1: increasing interaction strength softens the eigenvalues while leaving the eigenvectors, equilibrium abundances, and their noise exposure unchanged. Consequently, the critical-mode localization and accompanying changes in noise visibility that arise along our natural routes to instability are absent by construction.

More generally, our result provides the converse of a familiar
ambiguity in statistical signatures of criticality. Several
statistical signatures commonly associated with criticality can
arise without fine tuning to an underlying critical point
\cite{SchwabNemenmanMehta,MorrellSederbergNemenman,TouboulDestexhe}.
Conversely, we find that a genuine approach to a stability boundary
need not generate the expected pairwise critical signature. The
mapping is therefore non-invertible in both directions: apparent
critical statistics need not imply critical dynamics, and critical
dynamics need not imply extreme pairwise statistics. In ecological
communities this ambiguity is further compounded because equal-time
abundance correlations depend not only on deterministic interactions
but also on the covariance structure of environmental forcing
\cite{camacho2024nonequilibrium,goldberg2026niche,cholsky}.
Approaching criticality does not remove this dependence; it
selectively amplifies the components of the forcing that reach the
soft directions.

Finally, the limitation established here concerns equal-time
statistics, not the dynamics of critical slowing down itself. A mode
with small $\mu_1$ remains slow even when its instantaneous
covariance is masked by other modes, so information about
marginality should persist in relaxation times, low-frequency
fluctuations, and lagged correlations; indeed, spectral and
coherence-based measures, which exploit precisely this temporal
information, have recently been shown to recover ecological
structure that simple equal-time correlations miss
\cite{chen2025inferring}. 

Finite observation windows introduce an
additional caution, which is a direct consequence of the present
analysis: if the sampling window is shorter than the longest
relaxation time $1/\mu_1$, the slow mode is not adequately averaged
and can itself generate spuriously extreme sample correlations on
the species that carry it. Relatedly, Aguilar
\emph{et al.}~\cite{aguilar2026unraveling} showed that inferred
pairwise interaction measures depend on the duration and design of
the measurement protocol, to the point of sign reversals with no
change in the underlying ecological roles. Both effects point to the
same practical conclusion: correlation-based evidence near a
transition should be interpreted relative to the observation time as
well as to the intrinsic relaxation time (see further discussion in Supplementary
Material~\ref{app:fragility}). Criticality is a dynamical property of the community, not a universal
pattern in its instantaneous correlations.

{\bf Acknowledgments:} N.M.S. acknowledges support from the Israel Ministry of Science (Italy-Israel cooperation, grant no. 7578) and of the Israel Science Foundation (grant no. 2435/24).
\pagebreak

\bibliographystyle{unsrtnat}
\bibliography{ref}

\makeatletter
\newcounter{suppsec}
\renewcommand{\thesuppsec}{\Alph{suppsec}}
\newcommand{\suppsection}[2]{%
 \refstepcounter{suppsec}%
 \setcounter{equation}{0}\setcounter{figure}{0}\setcounter{table}{0}%
 \renewcommand{\theequation}{\thesuppsec\arabic{equation}}%
 \renewcommand{\thefigure}{\thesuppsec\arabic{figure}}%
 \renewcommand{\thetable}{\thesuppsec\arabic{table}}%
 \renewcommand{\thesubsection}{\thesuppsec\arabic{subsection}}%
 \renewcommand{\theHsubsection}{supp.\thesuppsec.\arabic{subsection}}%
 \setcounter{subsection}{0}%
 \label{#1}%
 \vspace{2.4ex}%
 \noindent{\sffamily\bfseries\large\color{black!88}%
 \thesuppsec.\enspace #2\par}%
 \vspace{0.8ex}%
}
\makeatother

\clearpage
\onecolumngrid
\begin{center}
{\sffamily\bfseries\LARGE Supplemental Material\par}
\end{center}
\vspace{0.25em}
\noindent{\color{RuleGray}\rule{\textwidth}{0.6pt}}
\vspace{0.5em}

\suppsection{app:engineered}{Construction of the engineered community}

The engineered community of Fig.~\ref{fig:naive} is designed to
realize the single-mode limit inside a fully legitimate competitive
community: all interactions positive, unit self-interaction, positive
definite interaction matrix, and a feasible equilibrium with equal abundances.
The construction is spectral. Let $\mathbf e=\mathbbm{1}/\sqrt S$ be the
uniform direction and draw a soft direction as
\begin{equation} \label{A1}
u_i \propto s_i\,e^{\zeta_i},\qquad
s_i=\pm1 \text{ (half each)},\qquad
\zeta_i\sim\mathcal N(0,h^2),
\end{equation}
projected orthogonal to $\mathbf e$ and normalized. A random
orthonormal basis $Q=(\mathbf e,\mathbf u,\mathbf q_3,\ldots,
\mathbf q_S)$ is completed by Gram--Schmidt, and we set
\begin{equation}
\alpha_0=Q\,\mathrm{diag}
\bigl(\tfrac S2,\;\mu_1,\;\lambda_3,\ldots,\lambda_S\bigr)\,
Q^{\mathsf T},
\qquad \lambda_a=e^{\xi_a},\ \ \xi_a\sim\mathcal N(0,h^2),
\end{equation}
followed by the diagonal normalization
$\alpha=D^{-1/2}\alpha_0 D^{-1/2}$ with $D=\mathrm{diag}(\alpha_0)$,
which enforces $\alpha_{ii}=1$ exactly.

Three properties hold simultaneously. First, positivity of all
off-diagonal entries is guaranteed by the stiff uniform mode: the
rank-one term $\tfrac S2\,\mathbf e\mathbf e^{\mathsf T}$ contributes
$1/2$ to every entry, while the remaining modes contribute
fluctuations of typical size $\sqrt{\mathrm{Var}(\lambda)/S}$, so all
entries stay positive as long as the spectral heterogeneity is not
too large. The heterogeneity $h=0.55$ used in Fig.~\ref{fig:naive} is
the largest value for which positivity holds for the realization
shown (off-diagonal entries span $0.04$--$0.52$). This heterogeneity is what spreads the
soft-mode weights and produces genuine distributions rather than
degenerate spikes in Fig.~\ref{fig:naive}(b)--(d); it also reduces
the participation number from $k=S$ to $k\simeq S/2$.
Second, $\alpha$ is positive definite by construction, since the
diagonal normalization preserves signature. Third, because
$\mathbf u\perp\mathbbm{1}$ and the same holds for every non-uniform
mode, $\alpha_0\mathbbm{1}=\tfrac S2\,\mathbbm{1}$, and after
normalization $\alpha\mathbbm{1}\propto\mathbbm{1}$ still holds up to the
small spread of $D$; choosing $r_i=(\alpha\mathbbm{1})_i$ makes
$\mathbf n^*=\mathbbm{1}$ an exact equilibrium. With equal abundances
$K=\alpha$, so the engineered spectrum of $\alpha$ is the relaxation
spectrum itself, and the Perron--Frobenius theorem makes the
mixed-sign character of the soft mode necessary rather than
artificial: in any all-positive competition matrix the uniform-sign
direction carries the \emph{largest} eigenvalue, so every soft
direction of such a community is mixed-sign, half the species rising
while half fall.

Only $\mu_1$ is varied between the panels of
Fig.~\ref{fig:naive}(b)--(d); the seed, and hence the gray spectrum,
the soft direction, and the background $\mathcal B=1.76$, are held
fixed. The visibility crossover marked in Fig.~\ref{fig:naive}(a) is
$\mu_1^\times=1/(k\mathcal B)=0.019$. The three panels sit at
$\mathcal V=49,\ 0.9,\ 0.11$, with
$\rho_{\rm rms}=0.90,\ 0.26,\ 0.07$ and strong correlations fractions
$f_{0.9}=0.63,\ 0,\ 0$.

\clearpage
\suppsection{app:models}{The three generative models}

In what follows we describe the types of realistic models of ecological communities analyzed through this paper. 

\subsection{A synthetic community} 

In a  synthetic community, such as that studied experimentally by \citet{crocker2025timescale}, a defined set of microbial isolates is deliberately combined in vitro, rather than obtained by sampling an intact naturally assembled community.

To model such a dynamics we implemented an interaction matrix whose off-diagonal terms  are drawn as
\begin{equation}
\alpha_{ij}=\bar\alpha+\frac{0.93\,\sigma_c}{\sqrt2}(A_{ij}+A_{ji}),
\qquad A_{ij}\sim\mathcal N(0,1),
\end{equation}
with $\bar\alpha=0.5$, $S=25$, $\sigma_c=0.05$. Again, $ \alpha_{ii}=1$. A draw is accepted if
$\mathbf n^*=\alpha^{-1}\mathbbm{1}$ is componentwise positive and
$\alpha$ is positive definite (acceptance rate around $0.13$); the main ensemble contains $600$ accepted communities. The parameters place the
accepted communities near the joint feasibility-stability edge.

For the size controls of Fig.~\ref{fig:corrs}(a) the correct protocol
is to match the \emph{distance to the edge}, not the interaction
statistics. Holding the May parameter $\sigma_c\sqrt S$ fixed does
not achieve this: the acceptance rate at fixed
$0.93\,\sigma_c\sqrt S=0.233$ collapses from $13\%$ at $S=25$ to
$7\%$ at $S=32$ and to $0.06\%$ at $S=75$, i.e.\ the joint
feasibility--stability constraint sharpens with size. We therefore
calibrate $\sigma_c(S)$ so that the median $\mu_s/\mu_{\rm med}$ of
the accepted ensemble matches the $S=25$ value of $0.142$, giving
$\sigma_c=0.044$ at $S=32$, $\sigma_c=0.025$ at $S=75$ and
$\sigma_c=0.0237$ at $S=80$ ($300$
communities each, realized medians $0.129$, $0.136$ and $0.143$). The
calibration must be done on the pair-capable mode: the same three
ensembles differ by up to $37\%$ in median $\mu_1/\mu_{\rm med}$, since the
softest mode is frequently localized on a single species and its position
carries no information about pair correlations. The dashed reference lines in
Fig.~\ref{fig:corrs}(a) are the medians of $\max_{i<j}|\rho_{ij}|$ of
these matched ensembles, the $S=32$ control matching the assembled
communities and the $S=80$ control the evolved ones.

\subsection{A local community assembled from a regional pool} 

A widely used model in the literature considers a  local community maintained by weak immigration from a regional species pool~\cite{kessler2015generalized,Bunin2017,barbier2018generic}. We implemented this model as follows:

A regional pool of $P=400$ species is drawn with mean competition  $\bar\alpha=0.8$ and
off-diagonal standard deviation $\sigma=0.02$. For such a community  the May-type
control parameter,
$X=(1-\bar\alpha)/\sqrt{\sigma^2P}$, equals $0.5$. This places the pool
well inside the multiple-equilibria regime~\cite{Bunin2017}, which we verify rather than
assume: on $97.6\%$ of the pools four random initial conditions relax to
more than one distinct clique, with median Jaccard distance $0.83$ between
them. The local community is assembled by
integrating the Lotka--Volterra dynamics with uniform growth rates, no
noise, and a small uniform
immigration $\lambda=10^{-8}$. The corresponding differential equation 
\begin{equation}
\dot x_i=x_i\Bigl(1-\sum_j\alpha_{ij}x_j\Bigr)+\lambda,
\end{equation}
was integrated from random initial conditions until the residual growth rate of the
residents falls below $10^{-3}$ and no absent species has a positive
invasion rate. We define the resident clique as the set of species for which $x_i>\sqrt\lambda$. 

The
interaction matrix restricted to the residents clique is then closed at
$\lambda=0$ by solving $\alpha_Q\mathbf n=\mathbbm{1}$ and iteratively
removing the most negative component. This step removes the one or two residents
that are in fact immigration-supported rather than self-sustaining
(median zero, at most three per community). One community is retained per
pool, the remaining initial conditions serving only to measure the
multiplicity quoted above. 

The ensemble studied here contains $250$
communities, all with positive definite restricted matrices, with median clique size
$Q=34$, median $\mu_s/\mu_{\rm med}=0.036$ and median
$\mu_1/\mu_{\rm med}=0.015$, a factor of four to five closer to the edge than the synthetic ensemble, with no parameter tuned to that end.

\subsection{Evolved communities}

 The evolutionary model follows
Ref.~\cite{shtilerman2015emergence} in the faithful parametrization
of the original implementation. A community of niche-overlap
competitors evolves by repeated speciation: a parent $p$ is chosen at random, a daughter is created with overlap
$\alpha_{dj}=h\,\alpha_{pj}+(1-h)\,\xi_j$ to every resident $j$
(inheritance $h=0.9$, innovation $\xi$ drawn with the pool
statistics), self-interaction $1$, and overlap
$\alpha_{dp}$ close to unity with its parent. After each speciation
the Lotka--Volterra equilibrium is recomputed  and species falling below the demographic cutoff
$10^{-5}$ are removed. 

Four replicate histories were run for $10^{8}$ speciation events each,
with the full interaction matrix and abundance vector stored every
$5\times10^{5}$ events up to $5.6\times10^{7}$ and every $10^{6}$ events
thereafter.
Communities used in the ensembles are snapshots with
$\geq2\times10^{7}$ events, past the initial diversification
transient; over this range the richness grows logarithmically in
evolutionary time. All spectral summaries used here
were statistically stationary,
so snapshots may be pooled. The ensemble contains $467$
communities with median $\mu_s/\mu_{\rm med}=0.032$ and median
$\mu_1/\mu_{\rm med}=0.016$. Recomputing the $\lambda=0$ equilibrium from
the stored matrix removes a median of one species per community,
consistent with the demographic cutoff used in the simulation.

Successive stored communities are separated by extensive turnover. 
At a separation of $5\times10^{5}$ speciation attempts, the median fraction 
of resident species replaced is $46\%$, and only $48.7\%$ of species pairs 
persist between successive snapshots.

\clearpage
\suppsection{app:numbers}{Numerical values behind the figures}
\subsection{The pair-capable soft mode $\mu_s$}

 All ensemble
statistics of the main text refer to the softest mode of
$K=N^{1/2}\alpha N^{1/2}$ that carries at least $20\%$ of its weight
off its dominant species, $1-\max_i u_a(i)^2\geq0.2$: a mode
localized on a single species cannot correlate any pair, and this
criterion excludes it without excluding the strongly localized but
two-species modes that do carry signal. 

The distinction matters
quantitatively: the pair-capable mode is not the softest mode in $53\%$,
$54\%$ and $53\%$ of the synthetic, assembled and evolved communities
respectively, and its median participation number is close to three in all
three ensembles, against $k=30$ for the engineered community of
Fig.~\ref{fig:naive}.  

For the selected mode $s$ we
record its rank in the spectrum, its participation number $k$, its
best pair $(i,j)=\arg\max|u_s(i)u_s(j)|$, the excitation
$A_s=\nu_s/\mu_s$, the background
$\mathcal B=S^{-1}\sum_{a\neq s}\nu_a/\mu_a$, the prediction of
Eq.~(\ref{eq:rhoEstimate}) evaluated for that pair with its exact
variances, and the measured $|\rho_{ij}|$ of the same pair from the full covariance.

\subsection{Figure ~\ref{fig:naive}} \label{C1}
\medskip

The figure is a controlled experiment on this construction: a single
community is built once, and only the soft eigenvalue $\mu_1$ is
dialed down across three decades, with the eigenvectors, the
background spectrum, and the noise held fixed. Every quantity quoted
in the caption is defined as follows.

\medskip  \noindent $S=60$ is the community size; all $S(S-1)/2=1770$ pairs enter the
histograms.
\medskip 

\noindent $h=0.55$ is the heterogeneity of the imposed soft direction and of
the background eigenvalues (the parameter of the log-normal draws in
the construction above, Eq. {\ref{A1}}). It is set at the largest value for which
all off-diagonal entries of $\alpha$ remain positive, so that the
community stays purely competitive; larger $h$ would produce
facilitative entries.
\medskip 

\noindent
Because $n_i^{*}=1$ for every species, the abundance weighting drops
out: the damping matrix coincides with $\alpha$, and with the
idiosyncratic forcing used here, $C^{(\eta)}=I$, every mode receives
unit exposure, $\nu_a=1$. All correlations are exact stationary
solutions of the covariance (Lyapunov) equation, obtained by
diagonalization.
\medskip 

\noindent
$k=1/\sum_i u_1(i)^4\simeq30$ is the participation number of the
soft eigenvector, the effective number of species that carry it;
by construction $k\simeq S/2$, a collective mode.
\medskip 

\noindent
$\mathcal B=\frac1S\sum_{a>1}\nu_a/\mu_a=1.76$ is the background
level: the mean per-species variance contributed by the $S-1$ modes
other than the soft one. It is the quantity against which the soft
mode must compete to be seen, and it does not change along the scan.
\medskip 

\noindent
With $\nu_1=1$ the visibility reduces to
$\mathcal V=1/(k\mu_1\mathcal B)$, so the crossover $\mathcal V=1$
sits at $\mu_1^{\times}=1/(k\mathcal B)=0.019$, the dashed line of
panel (a).
\medskip 

\noindent
Panels (b)--(d) use $\mu_1=3.8\times10^{-4}$, $2.1\times10^{-2}$ and
$1.9\times10^{-1}$ (realized values, after the diagonal
normalization), giving $\mathcal V=49$, $0.9$ and $0.11$: deep in
the visible regime, at the crossover, and in the hidden regime.
\medskip 

\noindent
$\rho_{\rm rms}$ is the root-mean-square of $\rho_{ij}$ over all
pairs ($0.90$, $0.26$, $0.07$), and $f_{0.9}$ is the fraction of
pairs with $|\rho_{ij}|>0.9$, the ``ears'' ($0.63$, $0$, $0$). Note
that in the hidden case the soft mode is still the slowest direction
in the community by a comfortable margin; what has changed is not
its softness but its visibility.
\medskip 

\noindent

\subsection{Figure~\ref{fig:scenarios}(d)--(f).}  \label{C2}

\medskip
\noindent
Medians with interquartile ranges over the full ensembles
($600$ synthetic, $250$ assembled, $467$ evolved communities). Each
entry reads median $[\mathrm{Q1},\mathrm{Q3}]$, where
$[\mathrm{Q1},\mathrm{Q3}]$ is the interquartile range, the interval
between the 25th and 75th percentiles that contains the central half
of the ensemble. We use medians and quartiles rather than means and
standard deviations because the underlying distributions are
strongly asymmetric with heavy tails, for which moment-based
summaries are dominated by a few extreme communities. All
mode-resolved entries refer to the pair-capable mode $s$ as explained above. Numbers are given in the following table:

\begin{center}
\renewcommand{\arraystretch}{1.3}
\begin{tabular}{@{}l@{\qquad}ccc@{}}
\toprule
 & synthetic & assembled & evolved\\
\midrule
distance,\; $\mu_s/\mu_{\rm med}$
  & $0.142\;[0.101,\,0.184]$
  & $0.036\;[0.020,\,0.059]$
  & $0.032\;[0.021,\,0.048]$\\
softest mode,\; $\mu_1/\mu_{\rm med}$
  & $0.072\;[0.038,\,0.118]$
  & $0.015\;[0.006,\,0.026]$
  & $0.016\;[0.008,\,0.026]$\\
signal,\; $\nu_s/(k\mu_s)$
  & $1.9\;[1.4,\,2.7]$
  & $9.9\;[6.3,\,17.4]$
  & $13.2\;[9.8,\,18.6]$\\
background,\; $\mathcal B$
  & $2.30\;[2.25,\,2.37]$
  & $6.87\;[6.45,\,7.56]$
  & $10.4\;[10.0,\,10.7]$\\[2pt]
\textbf{visibility,\; $\bm{\mathcal V_s}$}
  & $\bm{0.84\;[0.62,\,1.15]}$
  & $\bm{1.44\;[0.95,\,2.34]}$
  & $\bm{1.28\;[0.96,\,1.82]}$\\
participation,\; $k$
  & $3.3\;[2.1,\,5.2]$
  & $2.9\;[2.0,\,4.7]$
  & $3.1\;[2.1,\,5.1]$\\
\bottomrule
\end{tabular}
\end{center}

\medskip
\noindent
The absolute levels of the signal and background rows are not
intrinsic to the three mechanisms. Re-running the assembled ensemble from a
pool with $\bar\alpha=0.5$ instead of $0.8$, at essentially unchanged
clique size ($36$ against $34$), lowers the signal by a factor of $2.8$ and
the background by a factor of $2.5$, leaving $\mathcal V_s$ within $13\%$
of its value. The ordering across the three columns therefore reflects the
mean interaction strength of each generative model as well as the mechanism
itself, whereas the visibility ratio is robust to it.

\subsection{Figure~\ref{fig:corrs}(a).} \label{C3}
\medskip
\noindent
The strongest pair in each community, $\max_{i<j}|\rho_{ij}|$
(median $[\mathrm{Q1},\mathrm{Q3}]$ as above):

\begin{center}
\renewcommand{\arraystretch}{1.3}
\begin{tabular}{@{}l@{\qquad}c@{\qquad}c@{}}
\toprule
ensemble & $S$ & $\max_{i<j}|\rho_{ij}|$\\
\midrule
synthetic            & $25$      & $0.383\;[0.346,\,0.433]$\\
assembled            & ${\sim}34$ & $0.515\;[0.437,\,0.609]$\\
evolved              & ${\sim}80$ & $0.508\;[0.443,\,0.570]$\\
\midrule
synthetic control    & $32$      & $0.374\;[0.340,\,0.424]$\\
synthetic control    & $80$& $0.221\;[0.207,\,0.237]$\\
\bottomrule
\end{tabular}
\end{center}

\medskip
\noindent
These pooled medians mix the three ensembles' different coverage of
spectral depth, so the organizational enhancement must be read at
matched depth.  The enhancement over
the size-matched control is  a factor $1.2$ for assembly and $2.1$ for
diversification. Both are significant against the synthetic ensemble while
assembled and evolved are indistinguishable from each other.

\medskip
\noindent
Interestingly, the size correction works against the effect rather than producing
it. At matched depth the synthetic control falls with size, $0.402$ at
$S=25$ to $0.394$ at $S=32$ to $0.223$ at $S=80$, because the fluctuation
background is spread over more modes. Comparing the evolved communities
with the small synthetic ensemble rather than with their own size-matched
control would therefore understate the architectural effect, giving $1.3$
in place of $2.1$.

\medskip
\noindent
Four communities, all assembled cliques and $1.6\%$ of that ensemble,
have $\max_{i<j}|\rho_{ij}|>0.9$; none of the $600$ synthetic, $467$
evolved or $900$ control communities does.  Near-perfectly correlated pairs are thus not absent
in principle. They appear exactly where the visibility budget predicts
them, and the visibility required is reached only in the extreme tail of
one of the three routes.

\subsection{Figure~\ref{fig:corrs}(b).} \revb{\label{C4}}

\medskip
\noindent
The ratio of the measured to the variance-only-predicted correlation
of the best pair, and its mode-by-mode origin
(median $[\mathrm{Q1},\mathrm{Q3}]$ as above):

\begin{center}
\renewcommand{\arraystretch}{1.3}
\begin{tabular}{@{}l@{\qquad}ccc@{}}
\toprule
 & synthetic & assembled & evolved\\
\midrule
$|\rho|_{\rm meas}/|\rho|_{\rm pred}$
  & $0.46\;[0.32,\,0.61]$
  & $0.65\;[0.40,\,0.83]$
  & $0.48\;[0.31,\,0.64]$\\
remaining spectrum, mode-diagonal
  & $-0.61$ & $-0.48$ & $-0.60$\\
cross terms
  & $+0.04$ & $+0.06$ & $+0.06$\\
\bottomrule
\end{tabular}
\end{center}

\medskip
\noindent
The origin of the factor is fixed by a direct decomposition of the
pair covariance,
\begin{equation}
2{\tilde \Sigma}_{ij}=A_s\,u_s(i)u_s(j)
+\sum_{a\neq s}A_a\,u_a(i)u_a(j)
+\text{(cross terms)},
\end{equation}
whose last two contributions are quoted in the table as median
ratios relative to the soft-mode term: the mode-diagonal
contribution of the remaining spectrum carries the opposite sign and
roughly half the magnitude, while the cross terms are negligible.

\medskip
\noindent
The cancellation follows from the orthogonality sum rule
$\sum_a u_a(i)u_a(j)=0$: the pair weights of all modes sum to zero,
so a large positive weight in one mode forces a compensating
negative aggregate in the rest. The measured factor
${\sim}2$ is the dynamical realization of this exact geometric
constraint, with the near-universal value of the ratio across
ensembles reflecting how evenly the compensating weight is spread
over the spectrum.

\subsection{Width of the correlation distribution.} 

\medskip
 In addition to the
extreme-value statistics above and to allow comparison with former works~\cite{Calvo2026,Ferreira2025,Chen2019,ChenDNB2012} we recorded for every community the
overall width of its pairwise-correlation distribution,
$\rho_{\rm rms}=\langle\rho_{ij}^2\rangle^{1/2}$ over all pairs.

\begin{figure}[!htbp]
\preprintincludegraphics[width=0.74\linewidth]{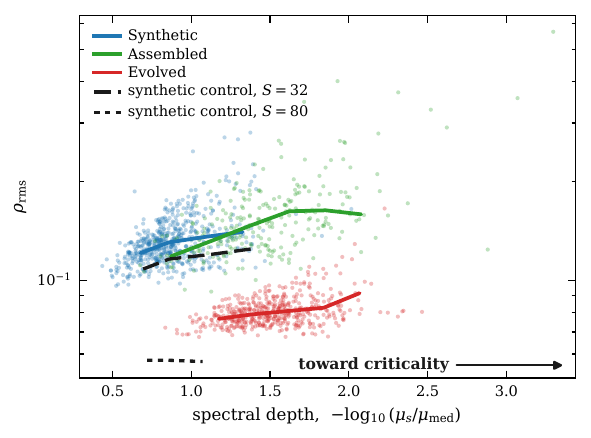}
\caption{\label{fig:rms}
Width of the pairwise-correlation distribution,
$\rho_{\rm rms}$, against the spectral depth of the pair-capable mode;
criticality is approached to the right. Points are individual communities
and heavy coloured lines binned medians, for the three natural ensembles.
The black dashed and dotted lines are the binned medians of the two
size-matched synthetic controls, $S=32$ and $S=80$. Within each ensemble
the width grows toward the edge. Across ensembles at fixed depth it varies
by a factor of about two, but this spread is dominated by community size
rather than by architecture: the evolved branch lies lowest of the three
while still sitting above its own $S=80$ control. }
\end{figure}

Figure~\ref{fig:rms} shows $\rho_{\rm rms}$ against the spectral
distance to criticality for the three natural ensembles, together
with the two size-matched synthetic controls. Within every ensemble
the width grows toward criticality, although moderately: the median
rises by $16\%$, $34\%$ and $20\%$ between the shallowest and the deepest
bin of the synthetic, assembled and evolved ensembles.

\medskip
\noindent
Since the estimator of Ref.~\cite{Calvo2026} is formulated in terms
of the pairwise-covariance rather than the Pearson-correlation
distribution, we repeated the analysis for the width of the
distribution of relative-abundance covariances,
$\Sigma_{ij}/(\bar n_i\bar n_j)$. The monotonic broadening toward
the edge persists and is stronger (second row of the table), so the
trend is not an artifact of the correlation normalization. The
architectural offset, however, is much larger in this observable: at
matched depth and size, the organized communities are broader than
their random controls by a factor of $8$--$11$ ($11.1$ for assembly
against its $S=32$ control and $7.7$ for diversification against its
$S=80$ control), against $1.1$--$1.3$ for the Pearson width, in the same
window and with $93$--$153$ communities per ensemble. The
excess resides in the amplitudes of the relative fluctuations rather
than in the pair geometry, consistent with the anomalous excitation
of the organized soft directions, and implies that a width-based
distance calibration is most architecture-sensitive precisely in the
covariance formulation.

\clearpage
\suppsection{app:fragility}{Practical conditions for detection}

 The fingerprints discussed in the
main text constrain where they can be observed. Three conditions
follow directly from the structure of the signal.

\begin{enumerate}
    \item {\bf Wide observation window.} The
equal-time covariance of the soft mode is a stationary property:
estimating it requires an observation window long compared with
$1/\mu_1$, or, equivalently, an ensemble of independent realizations
each older than $1/\mu_1$. A window (or realization age) short
compared with $1/\mu_1$ does not merely lose power; the unrelaxed
slow mode moves secularly through the window and manufactures
spuriously extreme sample correlations on its support, statistically
indistinguishable from genuine U-shaped correlation statistics. Any empirical report of extreme
correlations in a species-rich community should therefore be
conditioned on the window-to-relaxation-time ratio.

\item  {\bf Robustness against demographic stochasticity.}  The
carriers of the signal are the rarest species of the community, so
the sampling floor matters: count noise attenuates precisely the
pair correlations that carry the tilt, and presence thresholds can
remove the carriers altogether.

\item  {\bf No hidden driver.} Because a latent common
driver with loadings on rare species generates the same low-rank
cross-sectional covariance as a soft interaction mode, a single
cross-section cannot distinguish the two; discrimination requires
replication of the same pair set across independent cohorts, or
temporal information.

\end{enumerate}

 These conditions jointly explain why the
signatures analyzed here have not been reported in ecological
survey data, and delimit the observational designs, long
high-frequency series or many mature independent realizations, in
which they could be.

\end{document}